\documentclass[aps,prl,reprint,superscriptaddress]{revtex4-2}

\usepackage[utf8]{inputenc}
\usepackage{amsthm}
\usepackage{amsmath}
\usepackage{amsfonts}
\usepackage{color}
\usepackage{dsfont}
\usepackage{dsfont}
\usepackage{soul}
\usepackage{hyperref}
\usepackage{graphicx,xcolor}
\usepackage{braket}
\usepackage{comment}

\theoremstyle{definition}

\makeatletter
\newcommand*\bigcdot{\mathpalette\bigcdot@{.5}}
\newcommand*\bigcdot@[2]{\mathbin{\vcenter{\hbox{\scalebox{#2}{$\m@th#1\bullet$}}}}}
\makeatother

\theoremstyle{definition}
\begin{document}

\newcommand{\liege}{Institut de Physique Nucléaire, Atomique et de Spectroscopie, CESAM, Universit\'e de Li\`ege, 4000 Liège, Belgium}

\title{Symmetric Resourceful Steady States via Non-Markovian Dissipation}

\author{Baptiste Debecker}
\author{Eduardo Serrano-Ens\'astiga}
\author{Thierry Bastin}
\author{François Damanet}
\author{John Martin}
\affiliation{\liege}

\begin{abstract}
We prove a no-go theorem for symmetry-based dissipative engineering of collective-spin steady states: in spin-only Lindblad dynamics with jump operators linear in the collective-spin operators, any unique steady state exhibiting at least $\mathbb{Z}_2 \times \mathbb{Z}_2$ symmetry is necessarily the maximally mixed state. We then show that bath memory lifts this obstruction, enabling unique entangled steady states with a prescribed symmetry and a metrological gain, and providing a steady-state witness of non-Markovianity. Notably, this framework is largely insensitive to the microscopic details of the bath.

\end{abstract}

\date{\today}
\maketitle

\emph{Introduction}---Reservoir engineering has become a versatile route to preparing highly nonclassical quantum states~\cite{Poyatos1996Q, Verstraete2009Q, Plenio2002E, Kastoryano2011D, Lin2025W, Plenio1999C}. Of particular interest is the stabilization of a unique nontrivial steady state, which avoids the need for adiabatic state preparation and provides intrinsic robustness against initialization errors~\cite{DallaTorre2013D, Diehl2008Q, Fan2023C}. Recent work has further shown that the convergence toward such steady states can itself be accelerated~\cite{Pocklington2025A,Bao2025A}.

A natural strategy in dissipative state engineering is to exploit continuous or discrete symmetries~\cite{Diehl2010D,Leghtas2015C, Kimichi2016S, Yanay2018R, Li2025H, Young2025E}. For collective spins, rotational symmetry strongly constrains low-order moments, naturally selecting highly isotropic states. This is especially relevant for rotation metrology, since the symmetry of the state directly constrains the directional structure of the quantum Fisher information (QFI)~\cite{Goldberg2025R, Chryssomalakos2017O, Martin2020O, Hervas2025B, Goldberg2018Q, Piotrak2024P}. In the terminology of Ref.~\cite{Zimba2006A}, states with isotropic spin moments up to order $t$ are called $t$-anticoherent. Pure examples are highly non-classical~\cite{Baguette2017A}, and their metrological interest persists even when they are mixed~\cite{Ensastiga2025Q}.

\begin{figure}
    \centering
    \includegraphics[width=0.95\linewidth]{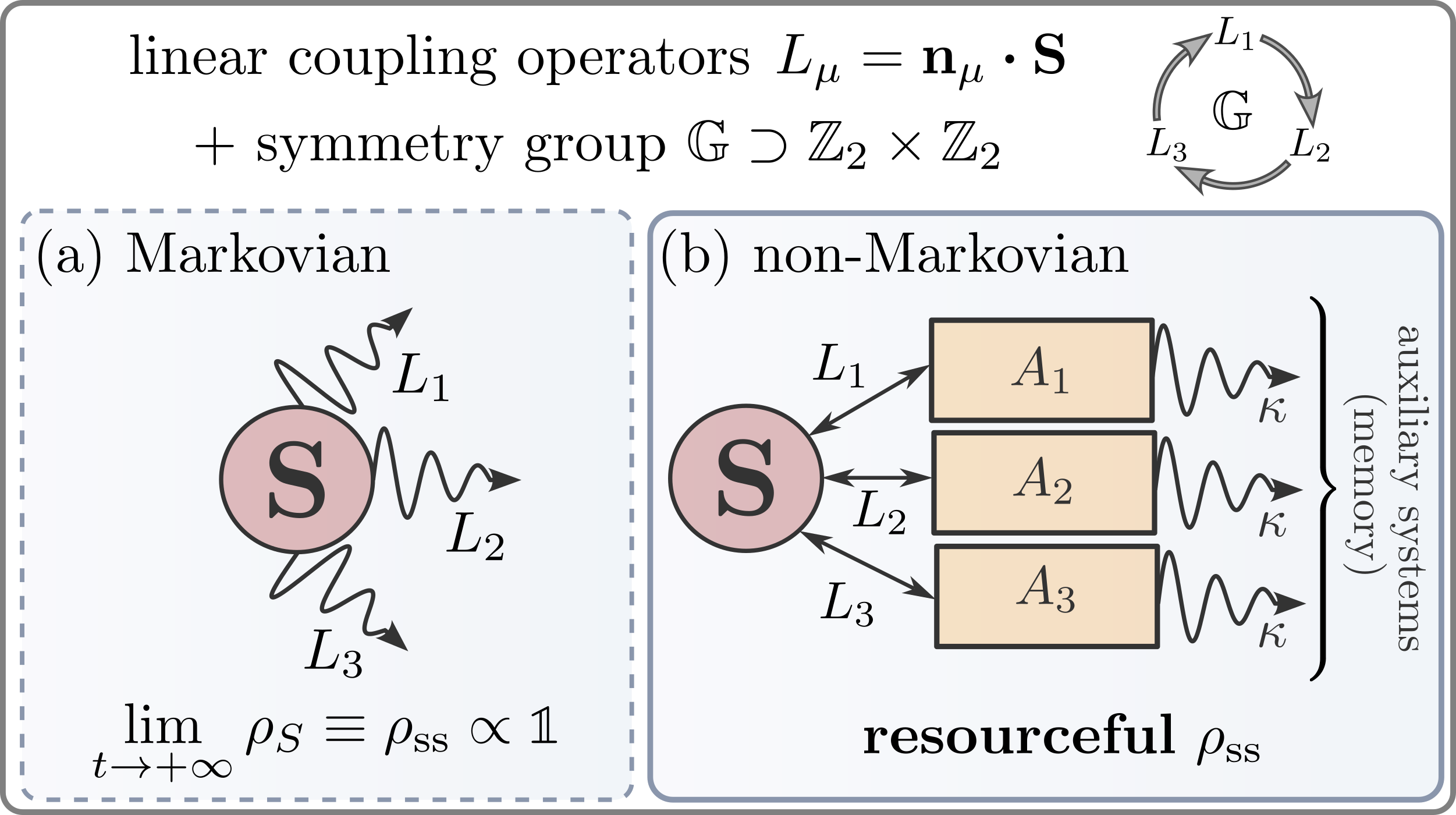}
    \caption{A collective spin $\mathbf{S}$ couples linearly to its environment through linear spin operators $L_\mu$ (three shown), chosen so that the set ${L_\mu}$ is closed under $\mathbb{G}$. (a) Spin-only Markovian dynamics: if the Lindbladian has weak symmetry containing $\mathbb{Z}_2\times\mathbb{Z}_2$, our no-go theorem implies that the MMS is the only possible unique steady state. (b) Non-Markovian case: the spin couples to damped auxiliary systems that provide bath memory.}
    \label{Fig1}
\end{figure}
In this Letter, we show that the dissipative engineering of collective-spin steady states with a prescribed symmetry is sharply limited in the Markovian regime. Specifically, we prove a no-go theorem: for any Lindblad generator with jump operators linear in collective spin $\mathbf{S}$, the presence of at least $\mathbb{Z}_2\times\mathbb{Z}_2$ weak symmetry (i.e., invariance under $\pi$-rotations about two orthogonal spin axes) forbids any unique steady state other than the maximally mixed state (MMS). Thus, within this broad and experimentally relevant class of spin-only Markovian models, symmetry alone cannot stabilize a unique nontrivial steady state.

We then show that bath memory provides a minimal and systematic route to circumvent this obstruction. Our approach is based on a Markovian embedding in which the collective spin couples linearly to a small set of auxiliary systems (AS), each itself damped by a Markovian reservoir (assumed to be at zero temperature, although this is not essential); see Fig.~\ref{Fig1}. The enlarged spin-plus-auxiliary dynamics is of Lindblad form, while the reduced spin dynamics is non-Markovian. Depending on the microscopic nature of the AS, this framework recovers bosonic pseudomodes~\cite{Imamoglu1994Stochastic, Dalton2001Theory, Garraway1997Nonperturbative, Pleasance2020Generalized, Debecker2024C}, fermionic mesoscopic-lead constructions~\cite{Brenes2020T, Chen2019M, Lacerda2024E}, or two-level fluctuator models relevant to gate errors~\cite{Papic2023F, Burnett2019D}.

By choosing the spin--AS coupling operators $L_\mu=\mathbf{n}_\mu\boldsymbol{\cdot}\mathbf{S}$ ($\mathbf{n}_\mu \in \mathbb{C}^3$) so that the set $\{L_\mu\}$ is closed under a target symmetry group, we construct unique reduced spin steady states that are both nontrivial and entangled. They realize genuinely non-Markovian steady states in the sense of Ref.~\cite{Ask2022N}. Conversely, our no-go theorem turns this observation into a simple witness of non-Markovianity based solely on the steady state.

Finally, we show that the resulting symmetry-engineered steady states are entangled and metrologically useful. Depending on the implemented symmetry, they provide either planar or fully isotropic quantum-enhanced sensitivity. The models we introduce for $\mathbb{Z}_2 \times\mathbb{Z}_2 $ (also denoted $\mathbb{D}_2$) and $\mathbb{U}(1)\times\mathbb{Z}_2$ symmetry are directly compatible with multimode cavity-QED architectures~\cite{Marsh2025M} and trapped-ion reservoir-engineering platforms~\cite{So2025Q, Sun2025Q}. On the other hand, recent cavity experiments have demonstrated controllable three- and four-body interactions~\cite{Luo2025R}, relevant to the higher polyhedral-symmetry constructions studied here.

\emph{Lindblad limit---} We consider $N$ spins-$\tfrac{1}{2}$ described by a collective spin $S_i = \frac12 \sum_{j=1}^N \sigma_i^{(j)}$ ($i=x, y, z$), written in terms of single-spin Pauli operators $\sigma_i^{(j)}$, and interacting with a generic environment. In the Markovian or Lindblad limit, the reduced spin state $\rho_S$ evolves according to
\begin{equation}
    \frac{d\rho_S}{dt} = -i [H_S, \rho_S] + \sum_{\mu} \frac{\gamma_\mu}{2} \mathcal{D}_{L_\mu}[\rho_S] \equiv \mathcal{L}[\rho_S],
    \label{eq::LindbladForm}
\end{equation}
with jump operators given by $L_\mu=\mathbf{n}_\mu\boldsymbol{\cdot}\mathbf{S}$, $\mathcal{L}$ the Lindbladian, and $\mathcal{D}_K[\boldsymbol{\cdot}]=2K \boldsymbol{\cdot} K^\dagger-\{K^\dagger K,\boldsymbol{\cdot}\}$.

Our goal is to prepare highly symmetric spin steady states $\rho_\mathrm{ss}$. We therefore consider weak symmetries~\cite{Buca2012} of the Lindbladian, namely any unitary superoperator $\mathcal{U}[\boldsymbol{\cdot}] = U \boldsymbol{\cdot} U^\dagger$  such that $[\mathcal{L}, \mathcal{U}] = 0$~\cite{Buca2012, Minganti2018Spectral}. An important consequence of the existence of a weak symmetry is that the steady state, if unique, must inherit the symmetry of the generator of the dynamics. Indeed, the steady state fulfills $\mathcal{L} [\rho_\mathrm{ss}] = 0$. Since $\mathcal{U}$ commutes with $\mathcal{L}$, it follows that $\mathcal{U} \mathcal{L}[\rho_\mathrm{ss}] = \mathcal{L}\,\mathcal{U}[\rho_\mathrm{ss}] = 0$. By uniqueness, we conclude that $\mathcal{U}[\rho_\mathrm{ss}] = \rho_ \mathrm{ss}$. This makes weak symmetry a natural tool for dissipative state engineering. However, as shown by the no-go theorem below, this strategy fails in the Lindblad limit.

\emph{Markovian no-go theorem}---Consider a spin-only Lindbladian $\mathcal{L}$ of the form~\eqref{eq::LindbladForm} and an arbitrary  Hamiltonian $H_S$. If $\mathcal{L}$ is at least $\mathbb{D}_2$-symmetric, i.e., $\mathcal{L}$ has two $\mathbb{Z}_2$-weak symmetries along perpendicular spin axes, then the MMS is necessarily a steady state. Furthermore, if at least two jump operators are linearly independent, then the MMS is the only steady state.

\emph{Sketch of proof}--- Uniqueness follows from Evans’ theorem~\cite{Evans1977I, Frigerio1978S, Spohn1976A, Spohn1980K}. The $\mathbb{D}_2$ symmetry severely constrains the allowed jump operators $L_\mu$, so that the set $\{L_\mu \}$ is unitarily equivalent to a set of Hermitian jump operators. Consequently, the MMS is a steady state. We provide a detailed proof in the Supplemental Material~\cite{SM}.

\emph{Non-Markovian engineering}---The no-go theorem can be circumvented in two conceptually distinct ways: (i) by engineering nonlinear jump operators in $\mathbf{S}$; or (ii) by retaining linear couplings while departing from the Markovian regime and exploiting bath memory. Here, we take the second route, and show that non-trivial steady states with a prescribed symmetry can be systematically engineered in a broad class of non-Markovian models.

The construction is as follows. We consider the spin system interacting with $N_E$ AS, each of which is itself damped by a Markovian bath at a rate $\kappa_\mu$, as shown for identical AS in Fig.~\ref{Fig1}. The master equation for the enlarged spin $+$ AS state $\rho$ is of Lindblad form, with a Hamiltonian given by 
\begin{equation}
   H = H_S + \sum_{\mu=1}^{N_E} \omega_\mu A_\mu^\dagger A_\mu +  \sum_{\mu=1}^{N_E} g_\mu (L_\mu A_\mu^\dagger + L_\mu^\dagger A_\mu),
   \label{eq::Htot}
\end{equation}
where $A_{\mu}$ denotes the lowering operator of the $\mu$-th AS (bosonic, fermionic, or spin), and
\begin{equation}
    \frac{d\rho}{dt} = \mathcal{L}_\mathrm{tot}[\rho] =  -i [H, \rho] + \sum_{\mu =1}^{N_E} \kappa_\mu\mathcal{D}_{A_\mu}[\rho].
    \label{eq::LindbladFormEmbedding}
\end{equation}
To impose a given symmetry group $\mathbb{G}$, we choose a Hamiltonian $H_S$ with the desired symmetry and linear couplings such that the set $\{L_\mu\}$ is closed under the action of $\mathbb{G}$. That is, for every $U_l \in \mathbb{G}$, the transformed operator $U_l L_\mu U_l^\dagger $ coincides, up to a phase, with another element of the same set. Crucially, if the AS  are identical ($g_\mu = g$, $\omega_\mu = \omega$, $\kappa_\mu = \kappa$ for all $\mu$), then the total Lindbladian $\mathcal{L}_\mathrm{tot}$ necessarily satisfies $[\mathcal{L}_\mathrm{tot}, \mathcal{U}_l] = 0$, where $\mathcal{U}_l[\boldsymbol{\cdot}] = U_l \boldsymbol{\cdot} U_l^\dagger$, because the action of $\mathcal{U}_l$ can be compensated by a simultaneous permutation of identical AS, as illustrated in Fig.~\ref{Fig1}. Thus, even though the individual couplings $L_\mu$ need not be invariant, the full embedded dynamics respects the prescribed symmetry $\mathbb{G}$. As a result, any unique steady state of the embedded Lindbladian yields, after tracing out the AS, a spin steady state with the desired symmetry. We discuss the phase factor freedom in ~\cite{SM}.

\emph{Minimal example and non-Markovian witness---}
Although our construction generally requires the AS to be identical, this hypothesis can be relaxed in the special case of the $\mathbb{D}_2$ group. The reason is that the action of $\mathbb{D}_2$ leaves certain operators $L_\mu$ invariant up to an innocuous sign. As a result, one does not need to rely on the identical nature of the AS to enforce the symmetry in the Markovian embedding. Concretely, let $\mathbb{D}_2$ be generated by the superoperators $\mathcal{U}_i[\boldsymbol{\cdot}] = U_i \boldsymbol{\cdot} U_i^\dagger$ with $U_i = e^{i \pi S_i}$ ($i=x, z$), i.e., rotations of $\pi$ about the $x$ and $z$ axes. Importantly, the action of $\mathcal{U}_i$ ($i=x, z$) on $S_j$ ($j = x, y, z$) leaves $S_j$ invariant up to a sign; for example, $\mathcal{U}_x [S_y] = -S_y$ and $\mathcal{U}_x [S_x] = S_x$. Thus, by choosing the linear coupling operators $L_1 = S_x$ and $L_2 = S_y$, the group maps each $L_\mu$ to $\pm L_\mu$ rather than mixing different elements of the set $\{L_1, L_2\}$. As a result, the Liouvillian of the Markovian embedding [Eqs.~\eqref{eq::Htot}-\eqref{eq::LindbladFormEmbedding}] remains $\mathbb{D}_2$-symmetric even for non-identical AS. Thus, any unique steady state of the enlarged dynamics yields a spin steady state with $\mathbb{D}_2$-symmetry, which implies that $\braket{\mathbf{S}} = \mathbf{0}$, i.e., the first-order moments vanish and the state is 1-anticoherent. 

\begin{figure}
    \centering
    \includegraphics[width=\linewidth]{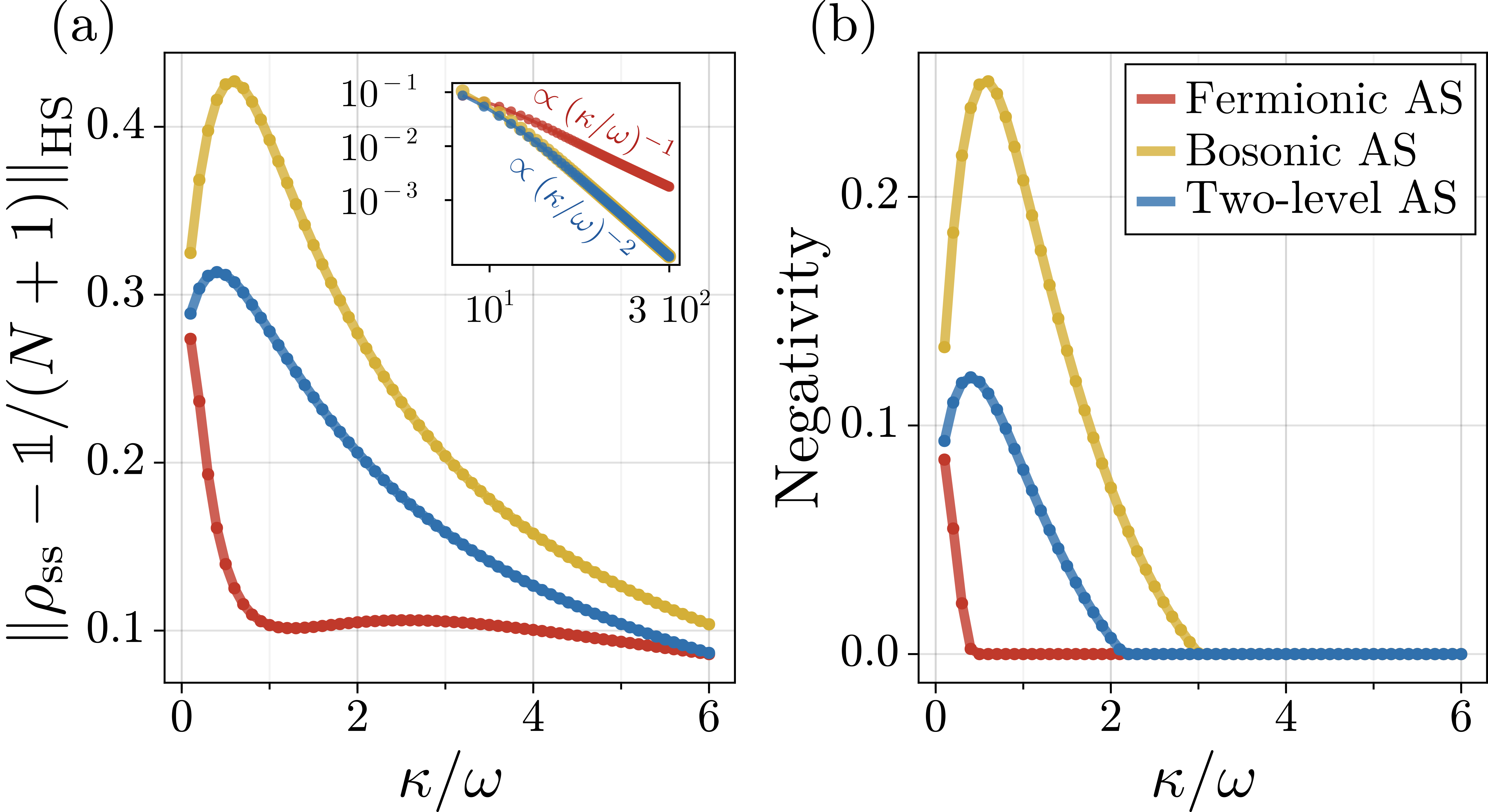}
    \caption{(a): Hilbert-Schmidt distance between the non-Markovian steady state and the MMS reached in the Lindblad limit, as a function of $\kappa/\omega$ for fermionic, bosonic and two-level AS. The inset shows that for fermionic AS the Lindblad limit is approached with a different scaling from bosonic and two-level AS. (b): Negativity for the $2|3$ bipartition as a function of $\kappa/\omega$. Here, $H_S = (h/N)S_z^2$, with $\omega_1 = \omega_2 = \omega$, $h/\omega=10$, $\gamma/\omega = 2.5$, $g_1 = g_2 =\sqrt{\gamma \kappa/{2N}} $, $N=5$ and $\kappa = \kappa_1 = \kappa_2$.}
    \label{fig2}
\end{figure}

A natural parametrization for the Hamiltonian \eqref{eq::Htot} is $g_1 = g_2 = \sqrt{\gamma \kappa /2N}$ with $\kappa_1 = \kappa_2 = \kappa$: in the limit $\kappa \rightarrow +\infty$ of ultra-fast damping, the AS can be eliminated exactly, and the reduced spin dynamics approaches a spin-only Lindblad description with linear jump operators $L_1 = S_x$ and $L_2 = S_y$~\cite{SM}. By virtue of the no-go theorem, its unique $\mathbb{D}_2$-symmetric steady state is the MMS. Figure~\ref{fig2}(a) quantifies the deviation from this Markovian limit through the excess purity relative to the MMS (or, equivalently, the Hilbert-Schmidt distance from the MMS). This quantity vanishes as $\kappa \rightarrow +\infty $, but remains finite for finite $\kappa$. Remarkably, this deviation is not merely a classical one: it comes with a finite negativity of $\rho_\mathrm{ss}$, as shown in Fig.~\ref{fig2}(b). This proves that the bath memory can stabilize nontrivial entangled $\mathbb{D}_2$-symmetric steady states using only coupling operators linear in $\mathbf{S}$. Conversely, this observation can be interpreted as a witness of non-Markovianity: within the class of $\mathbb{D}_2$-symmetric generators with jump operators linear in $\mathbf{S}$, any unique steady state observed experimentally that differs from the MMS certifies a departure from a spin-only Markovian Lindblad description.

While the symmetry argument is agnostic about the microscopic nature of the AS, the way in which the steady state departs from and returns to the Lindblad limit depends strongly on the structure of the Hilbert space of the AS. Figure~\ref{fig2} shows that, over a wide range of $\kappa/\omega$, bosonic pseudomodes drive the spin system further away from $\mathds{1}/(N+1)$ than two-level AS, consistent with the intuitive picture that bosonic modes provide a larger ``memory capacity" by virtue of their Hilbert space of infinite dimension. By contrast, bounded AS (two-level systems and fermions) generally produce smaller deviations because they can store less information. The fermionic case has an additional distinctive feature: because fermionic modes are represented via Jordan-Wigner strings, the second mode is dressed by the parity of the first mode, i.e. $A_2 = \sigma_z^{(1)} \sigma_-^{(2)}$, so that its effective action depends on the occupation of the first AS. In the formal limit $\kappa \rightarrow \infty $, the first mode is pinned to vacuum and the dressing reduces to an innocuous sign, thereby recovering the same Lindblad limit. At large $\kappa$, however, parity-dressed fermionic couplings lead to a distinct power-law convergence toward the Markovian fixed point, unlike bosonic and two-level AS, as highlighted in the inset of Fig.~\ref{fig2}(a) (see~\cite{SM} for a brief discussion of these exponents).

\emph{Continuous symmetry and metrology---}
Having introduced the relevant minimal symmetry group $\mathbb{D}_2$, we now demonstrate that our method can be used to engineer states with continuous symmetry, namely $\mathbb{U}(1) \times \mathbb{Z}_2$, which naturally leads to steady states that are useful in metrology. The continuous symmetry is generated by rotations around the $z$-axis, $\mathcal{U}_\phi[\boldsymbol{\cdot}] = U_\phi \boldsymbol{\cdot} U_\phi^\dagger$ with $U_\phi = e^{i\phi S_z}$, while the discrete symmetry is implemented by $\mathcal{U}_x$. A minimal realization is obtained by setting $H_S = (h/N) S_z^2$, $L_1 = S_-$, $L_2=S_+$, and identical AS. Indeed, the set $\{S_-, S_+\}$ is closed up to a phase factor under the group action since $\mathcal{U}_\phi[S_-] = e^{-i\phi}S_-,~\mathcal{U}_x[S_-] = S_+$, and thus fulfills the symmetry-closure condition of our construction. Consequently, the spin steady state necessarily inherits the continuous $\mathbb{U}(1) \times \mathbb{Z}_2$ symmetry, which translates into a diagonal form in the Dicke basis $\{\ket{N/2, m}\}$ ($m=-N/2, \ldots, N/2$) with symmetric populations $p_m = p_{-m}$. We stress that in the memoryless limit ($\kappa \rightarrow +\infty$), the no-go theorem applies, and the unique steady state is the MMS.

\begin{figure}
    \centering
\includegraphics[width=\linewidth]{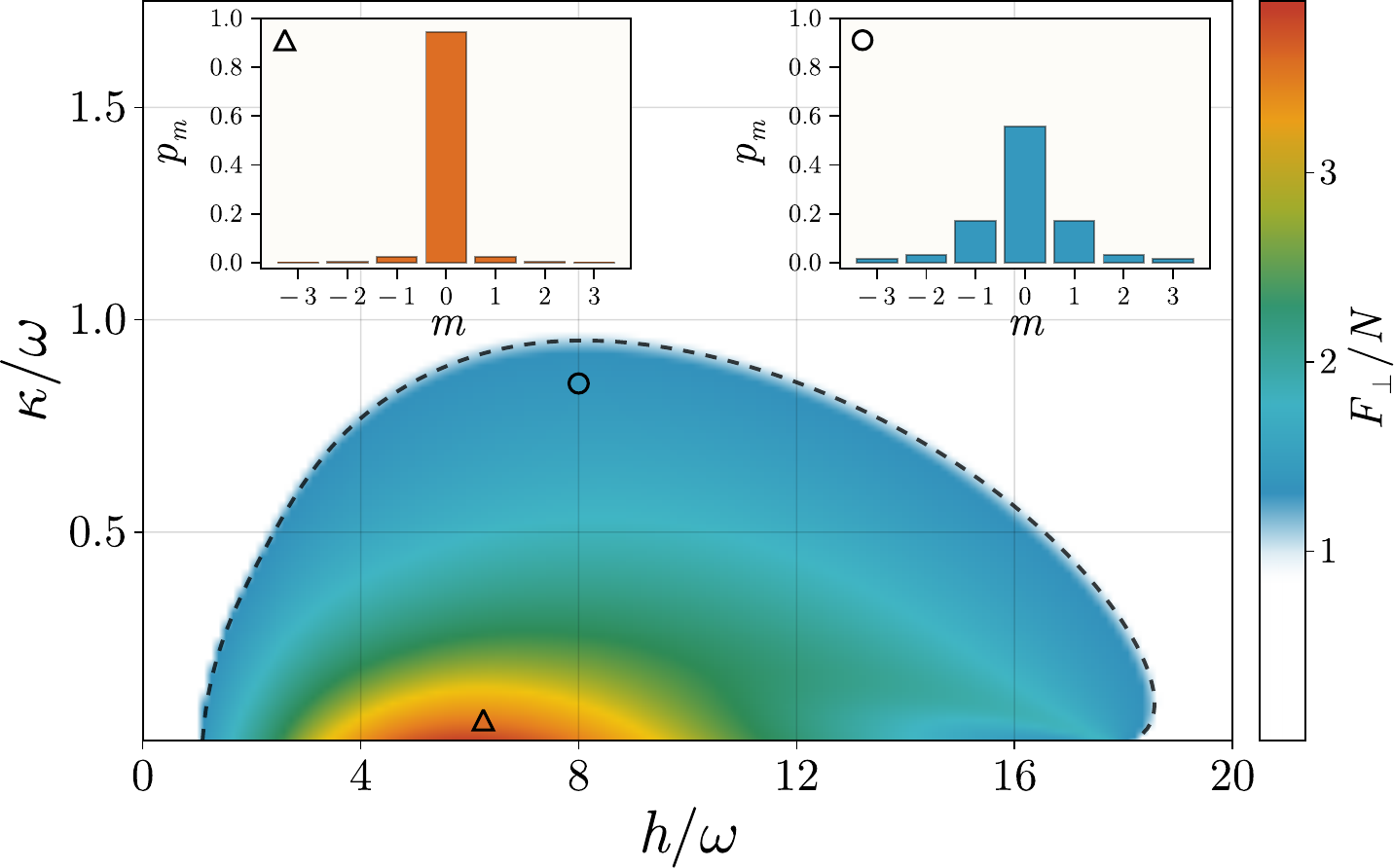}
    \caption{Equatorial QFI  for $H_S = (h/N) S_z^2$, $L_1 = S_-$ and $L_2 = S_+$ and with bosonic AS. The region of metrological gain ($F_\perp/N > 1$) is observed for $\kappa/\omega$ small enough; the white region corresponds to $F_\perp/N < 1$. The insets display the populations in the Dicke basis for the parameters indicated by the symbols (triangle and dot), showing a concentration around the state $m=0$ as the metrological gain increases. Other parameters: $\omega_1 = \omega_2 = \omega$, $g_1 = g_2 = \sqrt{\gamma \kappa/2N}$ with $\gamma = \omega$ and $N =6$. }
    \label{fig::3}
\end{figure}

To connect this symmetry to metrology, we evaluate the quantum Fisher information (QFI) $\mathcal{F}(\rho_\mathrm{ss}, A)$ of the steady state with respect to collective rotations generated by $A$. Because $\rho_\mathrm{ss}$ is diagonal in the $S_z$ eigenbasis, it is completely insensitive to rotations about $z$, and thus $\mathcal{F}(\rho_\mathrm{ss}, S_z) =0$. The metrological resource instead lies in the equatorial plane: for any unit vector $\mathbf{n}_\mathrm{eq} \perp \hat{\mathbf{z}}$, the symmetry forces an identical equatorial QFI, i.e., $\mathcal{F}(\rho_\mathrm{ss}, \mathbf{n}_\mathrm{eq}\cdot \mathbf{S}) \equiv F_\perp$. Importantly, separable states satisfy $F_\perp \leq N$~\cite{Hyllus2012F, Pezze2018Q}; $F_\perp > N$ therefore guarantees both multipartite entanglement and metrological gain relative to separable states. Figure~\ref{fig::3} displays the ``metrological phase diagram" $F_\perp/N$ in the $(h/\omega, \kappa/\omega)$ plane, with the region of gain ($F_\perp/N >1$) clearly delineated, for bosonic AS to highlight the close connection with cavity-QED models~\cite{Palacino2020, Moodie2018}. The inset shows representative Dicke-basis populations $p_m$ in the gain region, showing the symmetry $p_m = p_{-m}$ and a strong concentration near $m=0$. Crucially, the gain disappears as $\kappa/\omega$ increases, directly revealing the essential role played by the bath memory.

\emph{Higher polyhedral symmetries---}
Our Markovian-embedding construction provides a systematic route for stabilizing spin states with a prescribed symmetry, such as polyhedral rotation groups. The key difference from the $\mathbb{D}_2$ and $\mathbb{U}(1) \times \mathbb{Z}_2$ symmetries is that polyhedral symmetries constrain not only first-order moments but also moments of higher order, thereby enabling metrologically useful states without a preferred direction and imposing a higher degree of anticoherence (see~\cite{SM} for more details). 

Specifically, for discrete rotation groups such as the tetrahedral $\mathbb{T}$, octahedral $\mathbb{O}$, and icosahedral $\mathbb{I}$ groups, one can always find sets of operators $\{L_\mu = \mathbf{n}_\mu \boldsymbol{\cdot} \mathbf{S}\}$ that are closed under the action of $\mathbb{G}$, as well as a Hamiltonian with the required symmetry. In the SM~\cite{SM}, we provide a compact construction of the required axes $\{\mathbf{n}_\mu\}$ and the Hamiltonians for each group.

\begin{figure}
    \centering
\includegraphics[width=\linewidth]{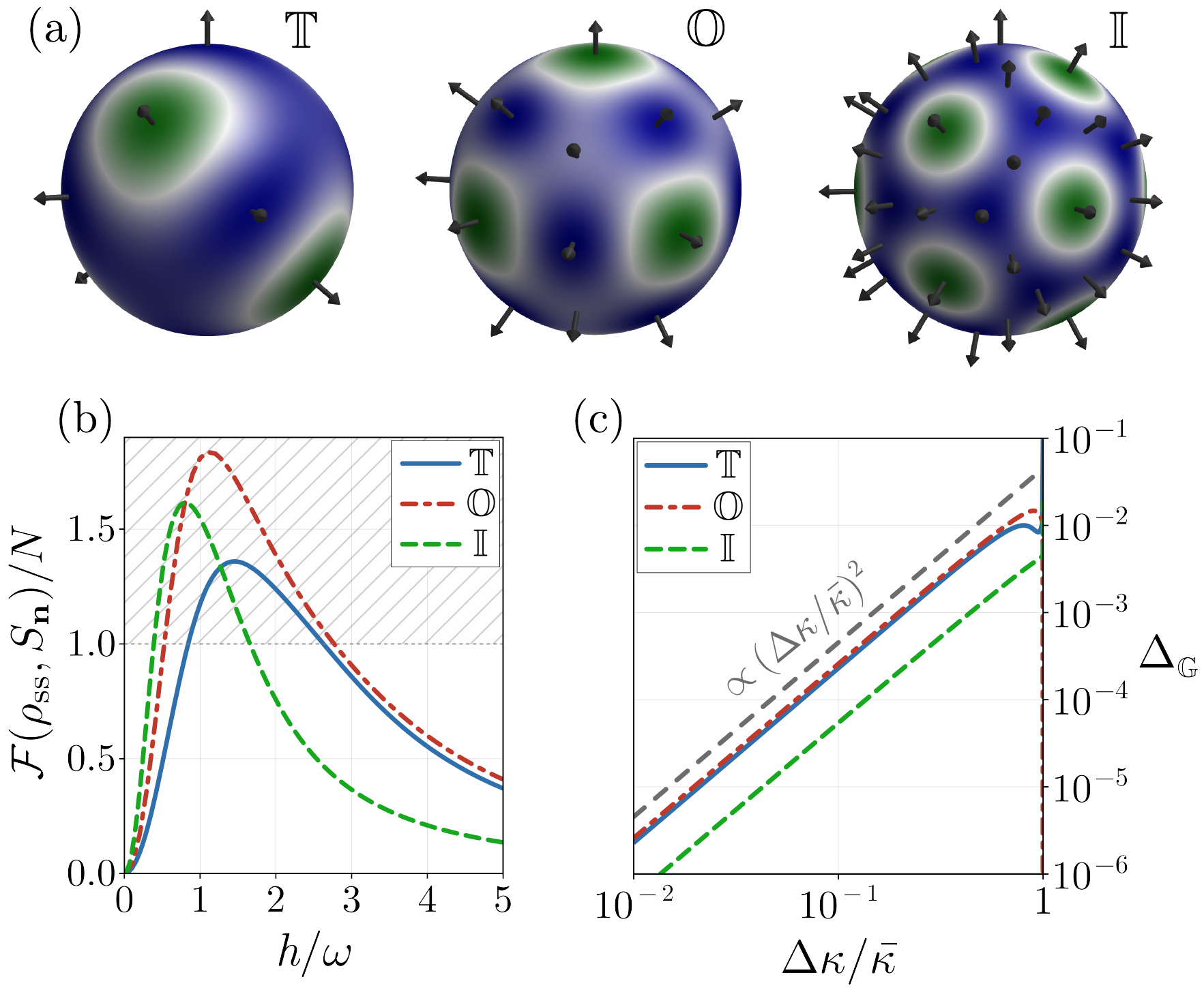}
    \caption{(a) Wigner functions of spin steady states for bosonic AS, showing polyhedral symmetries: tetrahedral ($N=4$), octahedral ($N=6$), and icosahedral ($N=6$). All rotational symmetry axes are shown. Colors from green to blue correspond to values from minimum to maximum. (b) Quantum Fisher information per spin-$\tfrac{1}{2}$, equal for all collective spin components $S_{\mathbf n}$ due to steady-state isotropy, as a function of the Hamiltonian strength $h/\omega$ (\cite{SM}for full models). The hatched region marks metrologically useful states.
(c) Symmetry-deviation parameter versus relative rate mismatch $\Delta\kappa/\overline{\kappa}$ (in log–log scale), showing quadratic scaling $ \Delta_\mathbb{G}\propto (\Delta\kappa/\overline{\kappa})^{2}$. Parameters: $\omega_k = \omega$ and $g_k = \sqrt{\gamma \kappa/2N}$ for all modes $k$, with $\gamma=\kappa = \omega/2$, and, for panel (c), $h=\omega$ for $\mathbb{T}$ and $\mathbb{O}$, and $h=\omega/6$ for $\mathbb{I}$.}
    \label{Fig:Wigner}
\end{figure}

Figure~\ref{Fig:Wigner} illustrates the obtained steady states and their metrological content. Panel~(a) shows the Wigner function for steady states symmetric under $\mathbb{T}, \mathbb{O}$ and $\mathbb{I}$, providing a visual representation of the implemented symmetry. Furthermore, the generated states necessarily have a fully isotropic QFI, i.e., $\mathcal{F}(\rho_\mathrm{ss}, \mathbf{n}\boldsymbol{\cdot} \mathbf{S})$ is independent of $\mathbf{n}$. As shown in panel (b), we find regions where this isotropic QFI exceeds the separable bound, resulting in quantum-enhanced metrological performance in all directions. These polyhedral steady states can be genuinely dissipative in origin. For example, when $N=5$, no pure $2$-anticoherent state exists. Therefore, a steady state with tetrahedral symmetry cannot be obtained from purely unitary dynamics alone. Our non-Markovian construction nevertheless stabilizes mixed, yet nontrivial, $2$-anticoherent states for $N=5$~\cite{SM}, highlighting an advantage of open-system symmetry engineering. Furthermore, the effect is robust. Panel (c) considers a spin coupled to three lossy bosonic modes with decay rates $\overline{\kappa}-\Delta\kappa$, $\overline{\kappa}$, and $\overline{\kappa}+\Delta\kappa$. A moderate imbalance $\Delta\kappa/\overline{\kappa}$ induces a moderate symmetry deviation
$\Delta_{\mathbb G}=1-\mathrm{Tr}(\rho_{\mathrm{ss}}\rho_{\mathbb G})/\mathrm{Tr}(\rho_{\mathrm{ss}}^2)$, where $\rho_{\mathbb G}=\frac{1}{|\mathbb G|}\sum_{l\in\mathbb G}U_l\rho_{\mathrm{ss}}U_l^\dagger$ is the group-averaged state. This deviation scales as $(\Delta\kappa/\overline{\kappa})^2$, which shows that the metrological gain does not require fine tuning of the decay rates. 

\emph{Conclusion---} 
We have identified a fundamental obstruction to the dissipative preparation of highly symmetric spin steady states within the Markovian paradigm: for spin-only, $\mathbb{D}_2$-symmetric Lindbladian with jump operators linear in $\mathbf{S}$, the only steady state that can be stabilized is the MMS. We then showed that non-Markovian generators admitting a general Markovian embedding do not suffer from this limitation: one can enforce the desired symmetry on the Markovian embedding and produce nontrivial and genuinely non-Markovian reduced spin steady states. Importantly, the construction relies on symmetry and is therefore insensitive to the microscopic nature of the AS. This leads to two practical results: (i) a witness of non-Markovianity based solely on the spin steady state and (ii) a systematic route to stabilizing resourceful states with prescribed continuous or discrete symmetries (including polyhedral rotation groups) using only linear-in-$\mathbf{S}$ coupling operators $L_\mu$. In particular, we showed the metrological usefulness in the equatorial plane for $\mathbb{U}(1) \times \mathbb{Z}_2$ and for all directions for higher polyhedral groups ($\mathbb{T}, \mathbb{O}$ and $\mathbb{I}$), as well as robustness against an imbalance in decay rates. Although we focused here on collective spin systems, the framework presented naturally extends to other many-body quantum systems with richer state spaces. More broadly,  this work establishes non-Markovianity as an enabling resource rather than a nuisance for steady-state engineering in spin platforms such as multimode cavity QED and trapped-ion reservoir engineering.

\begin{acknowledgments}
B.D. thanks C. Read for fruitful discussions. J.M., T.B., and E.S.-E.\ acknowledge the FWO and the F.R.S.-FNRS for their funding as part of the Excellence of Science program (EOS Project No.\ 40007526). T.B.\ also acknowledges financial support through IISN convention 4.4512.08. Computational resources were provided by the Consortium des Equipements de Calcul Intensif (CECI), funded by the Fonds de la Recherche Scientifique de Belgique (F.R.S.-FNRS) under Grant No.\ 2.5020.11.
\end{acknowledgments}

\bibliographystyle{apsrev4-2}
\bibliography{bib}

\end{document}


\newcommand{\liege}{Institut de Physique Nucléaire, Atomique et de Spectroscopie, CESAM, Universit\'e de Li\`ege, 4000 Liège, Belgium}

\title{Supplemental Material: Symmetric Resourceful Steady States via Non-Markovian Dissipation}
\author{Baptiste Debecker}
\affiliation{\liege}
\author{Eduardo Serrano-Ens\'astiga}
\affiliation{\liege}
\author{Thierry Bastin}
\affiliation{\liege}
\author{François Damanet}
\affiliation{\liege}
\author{John Martin}
\affiliation{\liege}

\maketitle

\setcounter{equation}{0}
\setcounter{figure}{0}
\setcounter{table}{0}
\setcounter{page}{1}
\makeatletter
\renewcommand{\theequation}{S\arabic{equation}}
\renewcommand{\thefigure}{S\arabic{figure}}
\renewcommand{\bibnumfmt}[1]{[S#1]}
\renewcommand{\citenumfont}[1]{S#1}

This Supplemental Material provides analytical and numerical details on the results presented in the main text of ``Symmetric Resourceful Steady States via Non-Markovian Dissipation''.

\section{Formal proof of the no-go theorem}
In this Section, we provide a formal proof of the no-go theorem stated in the main text. Without loss of generality, we consider $K$ jump operators $L_k = \alpha_k \,\mathbf{n}_k \boldsymbol{\cdot} \mathbf{S}$ that are traceless, since we can always make the transformation 
\begin{equation}
    \begin{aligned}
        \tilde{L}_k = & L_k + c_k
        \\
        \tilde{H}_S = &
        H_S - \frac{i}{2}  \sum_{k=1}^K \left( c_k^{*} L_k - \text{h.c.} \right)
    \end{aligned}
\end{equation}
that leaves the dynamics invariant and adds only terms to $H_S$ of the same order in the angular momentum operators. We focus on the dissipative part of the Lindblad master equation as the Hamiltonian part will not give additional constraints since $-i [H_S, \mathds{1}] = 0$ for any $H_S$. Without loss of generality, we assume that the two weak symmetries of the Lindbladian are given by $\mathcal{U}_x[\boldsymbol{\cdot}]= e^{-i\pi S_x} \,\boldsymbol{\cdot} \, e^{i \pi S_x}$ and $\mathcal{U}_y[\boldsymbol{\cdot}] = e^{-i\pi S_y} \,\boldsymbol{\cdot}\, e^{i \pi S_y}$, which send $(S_x,S_y,S_z)$ to $(S_x,-S_y,-S_z)$ and to $(-S_x,S_y,-S_z)$, respectively. Demanding that $\mathcal{U}_x$ and $\mathcal{U}_y$ are weak symmetries severely constrains the form of the jump operators. 

If $K=1$, then $L_1 \in \{\sqrt{\gamma_x}S_x, \sqrt{\gamma_y}S_y, \sqrt{\gamma_z}S_z\}$ comprises the only three jump operators satisfying the weak symmetries. Consequently, the MMS is a steady state because the jump operator is Hermitian. For $K=2$, we have two possibilities: either $L_1, L_2 \in \{\sqrt{\gamma_x}S_x, \sqrt{\gamma_y}S_y, \sqrt{\gamma_z}S_z\}$ or we have a special pair with equal rates $\{L_1, L_2\} = \{\sqrt{\gamma} S_+(n_1, n_2), \sqrt{\gamma}S_-(n_1, n_2)\}$ ($\gamma >0$). Here, $S_\pm (n_1, n_2) = c_1 S_{n_1} \pm c_2S_{n_2}$ with $n_1, n_2 \in \{x, y, z\}$ and $c_1, c_2$ complex constants. We stress that the equal rate condition cannot be relaxed here because one of the weak symmetries will necessarily transform $S_-(n_1, n_2)$ into $S_+(n_1, n_2)$ up to a phase factor and vice versa. In this case again the maximally mixed state is a steady state since either all jump operators are Hermitian, or $\sum_{i=1}^2 [L_i, L_i^\dagger] = \gamma [S_+(n_1, n_2), (S_+(n_1, n_2))^\dagger] + \gamma [S_-(n_1, n_2), (S_-(n_1, n_2))^\dagger] = 0 $. 

The last case to consider is $K = 3$. Indeed, for $K >3$, one can always perform a unitary transformation on the set of jump operators and return to the case where $K=3$, since  only three of them can be linearly independent. For completeness, we enumerate the two different routes that arise for $K = 3$. Either all jump operators are Hermitian and belong to $\{\sqrt{\gamma_x}S_x, \sqrt{\gamma_y}S_y, \sqrt{\gamma_z}S_z\}$, or there is one Hermitian operator $L_1 \in \{\sqrt{\gamma_x}S_x, \sqrt{\gamma_y}S_y, \sqrt{\gamma_z}S_z\}$ and a pair $\{S_+(n_1, n_2), S_-(n_1, n_2)\}$ with equal rates. 

Finally, we prove uniqueness as long as two linearly independent jump operators are considered. We do this by exploiting Evans theorem~\cite{Spohn1976A}, i.e., we show that the set of all operators which commute with all the jump operators, i.e. the commutant of the set of jump operators, is trivial. Namely, we want to show $\{L_1, L_2\}' = \{c\mathds{1}|c\in\mathbb{C}\}$. If both jump operators are Hermitian, we can safely assume $\{L_1, L_2\} = \{\sqrt{\gamma_x}S_x, \sqrt{\gamma_y}S_y\}$. In this case, if $O$ commutes with $L_1$ and $L_2$, it must also commute with $S_z$, hence the result. If we have a pair $\{S_+(n_1, n_2), S_-(n_1, n_2)\}$, then it is again easy to realize that any operator $O$ that commutes with $L_1$ and $L_2$ must also commute with $S_z$, which ends the proof.

\section{Phase factor degree of freedom}
In this Section, we formalize the phase degree of freedom mentioned and used in the main text. Namely, consider a group $\mathbb{G}$ and $U \in \mathbb{G}$ which acts on the spin system + AS of the form $U = U_S \otimes U_{AS}$ where $U_S$ and $U_{AS}$ act on the system Hilbert space and the AS respectively. We assume that the set $\{L_\mu\}$ is closed up to a phase under the action of the elements $U \in \mathbb{G}$, and we prove that a simple rotation of the AS can always compensate this phase factor. Specifically, we consider the master equation for the Markovian embedding
\begin{equation}
 \begin{split}
    \frac{d\rho}{dt} &= \mathcal{L}_\mathrm{tot}[\rho] =  -i [H, \rho] + \kappa \sum_{\mu} (2 A_\mu \rho A_\mu^\dagger - \{A_\mu^\dagger A_\mu, \rho\})
    \end{split}
    \label{eq::LindbladFormEmbedding}
\end{equation}
where 
\begin{equation}
   H = H_S + \omega \sum_{\mu=1}^{N_E}  A_\mu^\dagger A_\mu + g \sum_{\mu=1}^{N_E} (L_\mu A_\mu^\dagger + L_\mu^\dagger A_\mu).
   \label{eq::Htot}
\end{equation}
Now, the action of $\mathcal{U}[\boldsymbol{\cdot}] = U \boldsymbol{\cdot} U^\dagger $ on $H$ and the jump operators $A_\mu$ gives 
\begin{equation}
 \begin{split}
    \mathcal{U}[H] &= U_SH_S U_S^\dagger  + \omega \sum_{\mu=1}^{N_E}  U_{AS}A_\mu^\dagger A_\mu  U_{AS}^\dagger  + g \sum_{\mu=1}^{N_E} (U_S L_\mu U_S^\dagger U_{AS}A_\mu^\dagger U_{AS}^\dagger + U_S L_\mu^\dagger U_S^\dagger U_{AS} A_\mu U_{AS}^\dagger), \\
    \mathcal{U}[A_\mu] &= U_{AS}A_\mu U_{AS}^\dagger.
    \end{split}
\end{equation}
In our construction, the spin Hamiltonian is $U_S$-symmetric, which implies $U_S H_S U_S^\dagger = H_S$. Furthermore, by hypothesis there exists at least one operator $L_\alpha \in \{L_\mu\}$ such that   $U_S L_\alpha U_S^\dagger = e^{i \theta} L_\beta $ with $L_\beta \in \{L_\mu\}$ and $\theta \in \mathbb{R}$. Crucially, this phase factor can be compensated to keep the interaction term in \eqref{eq::Htot} invariant by choosing $U_{AS}$  such that
$U_{AS} A_\beta^\dagger U_{AS}^\dagger = e^{-i\theta} A_\beta^\dagger$, which is always possible since $e^{-i \theta A^\dagger_\beta A_\beta} A_\beta^\dagger e^{i\theta A^\dagger_\beta A_\beta} = e^{-i\theta} A_\beta^\dagger$, which follows from $[A^\dagger_\beta A_\beta, A^\dagger_\beta]= A^\dagger_\beta $ and the Baker-Campbell-Hausdorff lemma. Finally, because the Lindblad master equation is invariant if the jump operators are multiplied by a phase factor and the free auxiliary term $\sum_{\mu=1}^{N_E}  A_\mu^\dagger A_\mu$ is also unchanged under such phases, we conclude that the the potential phase factor appearing in the jump operators after the action of a symmetry is irrelevant.

\section{Memoryless limit $\kappa \rightarrow \infty$ and  adiabatic elimination of the Auxiliary Systems}

\subsection{Memoryless limit $\kappa \rightarrow \infty$}

In this section, we provide details about the memoryless limit $\kappa \rightarrow +\infty$ introduced in the main text restricting for simplicity to the special case of a unique bosonic AS. Recall that if the bath is quadratic, it can be formally integrated out. Consequently, the dynamics of the reduced spin system is entirely determined by the bath correlation function $\alpha(t-s)$ defined by 
\begin{equation}
    \alpha(t-s) =|g|^2 \braket{a(t) a^\dagger(s)},
\end{equation}
with $a(t)$ the bosonic annihilation operator evolved by the free part of the Lindbladian of the Markovian embedding, i.e., 
\begin{equation}
    \dot{a} = \mathcal{L}^\dagger [a] = i \omega [ a^\dagger a, a] + \kappa \mathcal{D}^\dagger [a] = (-i\omega-\kappa)a.
\end{equation}
Thus, if the residual Markovian bath to which the auxiliary system couples is at zero temperature, we obtain 
\begin{equation}
    \alpha(t-s) = |g|^2 e^{-i\omega (t-s) -\kappa (t-s)},\qquad t\geq s.
\end{equation}
Setting $g^2 = \gamma\kappa/2N$ and taking the limit $\kappa \rightarrow +\infty $ renders the correlation function trivial
\begin{equation}
    \alpha(t-s) = \frac{\gamma \kappa}{2N} e^{-i\omega (t-s)-\kappa (t-s) } \longrightarrow \frac{\gamma}{N} \delta(t-s)~\mathrm{if}~\kappa \rightarrow +\infty.
\end{equation}
Because the correlation function is the memory kernel determining the dynamics of the reduced system, $\alpha(t-s) \propto \delta(t-s)$ proves that the bath is strictly memoryless as only times $s=t$ will contribute to the reduced dynamics of the spin system. In this particular limit, the spin dynamics is then \emph{exactly} captured by the Lindblad master equation 

\begin{equation}
    \frac{d\rho}{dt} = -i [H_S, \rho] +  \frac{\gamma}{2N} (2L \rho L^\dagger - L^\dagger L \rho - \rho L^\dagger L ). 
\end{equation}
with $\rho$ the reduced density matrix of the spin degrees of freedom.

Note that a similar reasoning holds true for both i) non-interacting damped fermionic auxiliary systems provided that the auxiliary plus its residual bath remains quadratic, and ii) non-quadratic two-level auxiliaries, which in the limit of large $\kappa$ can be seen as an effective bosonic degree of freedom with at most one excitation. \\

\subsection{Adiabatic elimination of the auxiliary systems}

For large $\kappa$ one might expect a standard perturbative reduced description of the spin dynamics, obtained by adiabatically eliminating the AS modes, to correctly capture the steady states. Surprisingly, this is not the case, as discussed below.

Indeed, such a common second-order treatment is the Redfield master equation, which we write as follows~\cite{Link2022}
\begin{equation}
  \frac{d \rho}{dt} = -i[H_S, \rho] + \sum_\mu \left([\bar{L}_\mu \rho, L_\mu^\dagger] + [L_\mu, \rho\bar{L}_\mu^\dagger]\right),
  \label{eq::RedfieldME}
\end{equation}
where we defined 
\begin{equation}\label{rotatedjump}
 \begin{split}
  \bar{L}_\mu &= \int_0^{+\infty}dt~\alpha_\mu(t)L_\mu(t) \\
  \end{split}
\end{equation}
with $\alpha_\mu$ the correlation function of the $\mu^\mathrm{th}$ bath and $L_\mu$ the coupling operator mediating the interaction between the spin system and the bath $\mu$. For $\kappa \to \infty$, we have $\alpha(t) = \gamma/N \delta (t)$, and the Redfield master equation~(\ref{eq::RedfieldME}) reduces to the Lindblad spin-only master equation whose unique steady state is the MMS, as expected. For large but finite $\kappa$, however, the rotated jump operators~(\ref{rotatedjump}) become non-linear in terms of collective spin operators, thereby freeing the system from the constraints of the no-go theorem and making it possible to reach a steady state that departs from the MMS. Yet numerical simulations show that, despite this, the Redfield master equation~\eqref{eq::RedfieldME} cannot reproduce all the distinct asymptotic exponents with which the bosonic/two-level and fermionic embeddings approach the Lindblad limit. This indicates that the different exponents discussed in the main text originate from contributions \textit{beyond} second order, thereby highlighting the genuinely non-Markovian nature of our dissipative state-preparation scheme. 


\section{Hamiltonians and jump operators for the polyhedral groups} 

In this section, we explain the models considered in the main text for the tetrahedral ($\mathbb{T}$), octahedral ($\mathbb{O}$) and icosahedral ($\mathbb{I}$) symmetry groups.

\subsection{Tetrahedral symmetry}

A minimal Hamiltonian exhibiting tetrahedral ($\mathbb{T}$) symmetry is (see, e.g., Ref.~\cite{Read2025})
\begin{equation}
    \label{Htetra}
    H_{\mathrm{tetra}}
    =
    \frac{h}{N^2}
    \left[(S_+^2+S_-^2)S_z+S_z(S_+^2+S_-^2)\right].
\end{equation}

Another geometrically natural construction is based on the four $C_3$ symmetry axes of the tetrahedron,
\begin{equation}
    \mathbf{n}_1=
    \left(\sqrt{\frac{2}{3}},\,0,\,\frac{1}{\sqrt{3}}\right),\quad
    \mathbf{n}_2=
    \left(-\sqrt{\frac{2}{3}},\,0,\,\frac{1}{\sqrt{3}}\right),\quad
    \mathbf{n}_3=
    \left(0,\,\sqrt{\frac{2}{3}},\,-\frac{1}{\sqrt{3}}\right),\quad
    \mathbf{n}_4=
    \left(0,\,-\sqrt{\frac{2}{3}},\,-\frac{1}{\sqrt{3}}\right),
\end{equation}
through the Hamiltonian
\begin{equation}
    H'_{\mathrm{tetra}}
    =
    \frac{h}{N^2}\sum_{k=1}^4 (\mathbf{n}_k\boldsymbol{\cdot}\mathbf{S})^3.
\end{equation}

For the dissipative part, the smallest set of jump operators that is invariant under the action of $\mathbb{T}$ as a whole, and that yields a nontrivial steady state (that is, one different from the maximally mixed state), is
\begin{equation}
    \label{Ltetra}
    L_{\mathrm{tetra}}=\{S_x,S_y,S_z\}.
\end{equation}
An alternative choice is
\begin{equation}
    \label{Lptetra}
    L'_{\mathrm{tetra}}=\{\mathbf{n}_k\boldsymbol{\cdot}\mathbf{S}:k=1,2,3,4\}.
\end{equation}

Using either $L_{\mathrm{tetra}}$ or $L'_{\mathrm{tetra}}$, together with either $H_{\mathrm{tetra}}$ or $H'_{\mathrm{tetra}}$, leads to very similar nontrivial steady states. The results reported in Fig.~4 of the manuscript were obtained with HEOM~\cite{Tanimura89, Link2022, Debecker2024S} (setting the hierarchy depth $k_{\mathrm{max}}=5$), using the Hamiltonian in Eq.~\eqref{Htetra} and the jump operators in Eq.~\eqref{Ltetra}.

\subsection{Octahedral symmetry}

A minimal Hamiltonian exhibiting octahedral ($\mathbb{O}$) symmetry is
\begin{equation}
    \label{Hocta}
    H_{\mathrm{octa}}
    =
    \frac{h}{(N/2)^3}
    \left(S_x^4+S_y^4+S_z^4\right).
\end{equation}
This is the natural quartic invariant associated with the cubic/octahedral symmetry group.

For the dissipative part, a minimal set of jump operators leading to a nontrivial steady state is
\begin{equation}
    \label{Locta}
    L_{\mathrm{octa}}=\{S_x,S_y,S_z\}.
\end{equation}

Using $H_{\mathrm{octa}}$ together with $L_{\mathrm{octa}}$ yields a nontrivial steady state with octahedral symmetry (see Fig.~4). As in the tetrahedral case, the resulting steady state is highly isotropic and clearly differs from the maximally mixed state.

\subsection{Icosahedral symmetry}

A minimal Hamiltonian exhibiting icosahedral ($\mathbb{I}$) symmetry is
\begin{equation}
    \label{Hicosa}
    H_{\mathrm{icosa}}
    =
    \frac{h}{(N/2)^5}
    \sum_{k=1}^{6} (\mathbf{n}_k\boldsymbol{\cdot}\mathbf{S})^6,
\end{equation}
where the six symmetry directions are defined by
\begin{equation}
    \mathbf{n}_k=\frac{\tilde{\mathbf{n}}_k}{\sqrt{1+\phi^2}},
\end{equation}
with $\phi=(1+\sqrt{5})/2$ the golden ratio and
\begin{equation}
    \tilde{\mathbf{n}}_1=
    \left(0,1,\phi\right),\quad
    \tilde{\mathbf{n}}_2=
    \left(0,1,-\phi\right),\quad
    \tilde{\mathbf{n}}_3=
    \left(1,\phi,0\right),\quad
    \tilde{\mathbf{n}}_4=
    \left(1,-\phi,0\right),\quad
    \tilde{\mathbf{n}}_5=
    \left(\phi,0,1\right),\quad
    \tilde{\mathbf{n}}_6=
    \left(\phi,0,-1\right).
\end{equation}
These vectors specify six equivalent 5-fold symmetry axes of the icosahedron.

For the dissipative part, the minimal set of jump operators is again
\begin{equation}
    \label{Licosa}
    L_{\mathrm{icosa}}=\{S_x,S_y,S_z\}.
\end{equation}

Using $H_{\mathrm{icosa}}$ together with $L_{\mathrm{icosa}}$ yields a nontrivial steady state with icosahedral symmetry (see Fig.~4). Among the Platonic cases considered here, this construction provides the highest degree of discrete rotational symmetry.

\section{Symmetry and order of anticoherence}

In this section, we recall the notion of anticoherence and show how it follows directly from point-group symmetry for the cases considered in the main text.

Let $\rho$ be a spin-$s$ state. We use the irreducible tensor operators $T^{(s)}_{LM}$, with $L=0,\dots,2s$ and $M=-L,\dots,L$, as an operator basis. Following Ref.~\cite{Giraud2015}, a mixed state $\rho$ is said to be $t$-anticoherent if and only if
\begin{equation}
    \Tr\!\left(\rho\,T^{(s)}_{LM}\right)=0,
    \qquad
    L=1,\dots,t,\;\; M=-L,\dots,L.
\end{equation}
Equivalently, all multipole moments of rank up to $t$ vanish.

To connect this definition with symmetry, let $H_L$ be any operator belonging to the irreducible subspace of rank $L$, i.e. any linear combination of the operators $T^{(s)}_{LM}$ for fixed $L$. We define its group average over a finite symmetry group $\mathbb G$ as
\begin{equation}
    \overline{H}^{\,\mathbb G}_L
    \equiv \frac{1}{|\mathbb G|}
    \sum_{g\in\mathbb G} g\,H_L\,g^\dagger .
    \label{eq:group_average_HL}
\end{equation}
A key observation, proven in Ref.~\cite{Read2025}, is that the group average
\begin{equation}
    \overline{H}^{\,\mathbb G}_L=0
\end{equation}
vanishes for the low-rank sectors listed in Table~\ref{tab:group_average_vanishing}.

\begin{table}[h]
    \centering
    \begin{tabular}{c@{\hspace{1.5cm}}c}
        \hline\hline
        Symmetry group $\mathbb G$ & Vanishing ranks $L$ \\
        \hline
        $\mathbb{D}_2$ & $L=1$ \\
        $\mathbb T$ & $L=1,2$ \\
        $\mathbb O$ & $L=1,2,3$ \\
        $\mathbb I$ & $L=1,2,3,4,5$ \\
        \hline\hline
    \end{tabular}
    \caption{Ranks for which the group-averaged operator $\overline{H}^{\,\mathbb G}_L$ vanishes for the symmetry groups considered here.}
    \label{tab:group_average_vanishing}
\end{table}

We now use this result to determine the order of anticoherence implied by symmetry.

\begin{theorem}
A mixed spin state whose symmetry group contains $D_2$, $\mathbb T$, $\mathbb O$, or $\mathbb I$ is respectively $1$-, $2$-, $3$-, or $5$-anticoherent.
\end{theorem}

\begin{proof}
Any spin-$s$ density matrix can be expanded in the irreducible tensor basis as
\begin{equation}
    \rho
    =
    \sum_{L=0}^{2s} \boldsymbol{\rho}_L \boldsymbol{\cdot} \mathbf{T}_L,
    \label{eq:rho_tensor_expansion}
\end{equation}
where
\begin{equation}
    \mathbf{T}_L = \bigl(T^{(s)}_{L,-L},\dots,T^{(s)}_{L,L}\bigr),
\end{equation}
and
\begin{equation}
    \boldsymbol{\rho}_L = \bigl(\rho_{L,-L},\dots,\rho_{L,L}\bigr),
    \qquad
    \rho_{LM}=\Tr\!\left(\rho\,T_{LM}^{(s)\dagger}\right).
\end{equation}

Assume now that $\rho$ is invariant under the action of a symmetry group $\mathbb G$, that is,
\begin{equation}
    \overline{\rho}^{\,\mathbb G}=\rho,
\end{equation}
where $\overline{\rho}^{\,\mathbb G}$ denotes the average of $\rho$ over the group. Applying the group average to the expansion \eqref{eq:rho_tensor_expansion} gives
\begin{equation}
    \rho
    =
    \sum_{L=0}^{2s} \boldsymbol{\rho}_L \boldsymbol{\cdot} \overline{\mathbf{T}}^{\,\mathbb G}_L.
\end{equation}
By the observation above, the averaged irreducible tensor operators vanish up to a rank fixed by the group:
\begin{align}
    \mathbb G \supset \mathbb{D}_2 &\;\Rightarrow\; \boldsymbol{\rho}_L=0 \quad \text{for } L=1,\\
    \mathbb G \supset \mathbb T &\;\Rightarrow\; \boldsymbol{\rho}_L=0 \quad \text{for } L=1,2,\\
    \mathbb G \supset \mathbb O &\;\Rightarrow\; \boldsymbol{\rho}_L=0 \quad \text{for } L=1,2,3,\\
    \mathbb G \supset \mathbb I &\;\Rightarrow\; \boldsymbol{\rho}_L=0 \quad \text{for } L=1,\dots,5.
\end{align}
Therefore,
\begin{equation}
    \Tr\!\left(\rho\,T^{(s)}_{LM}\right)=0
\end{equation}
for all $M=-L,\dots,L$ and for all ranks $L$ up to $1$, $2$, $3$, or $5$, respectively. By definition, this means that $\rho$ is $1$-, $2$-, $3$-, or $5$-anticoherent, completing the proof.
\end{proof}

\begin{figure}
    \centering
    \includegraphics[width=0.5\linewidth]{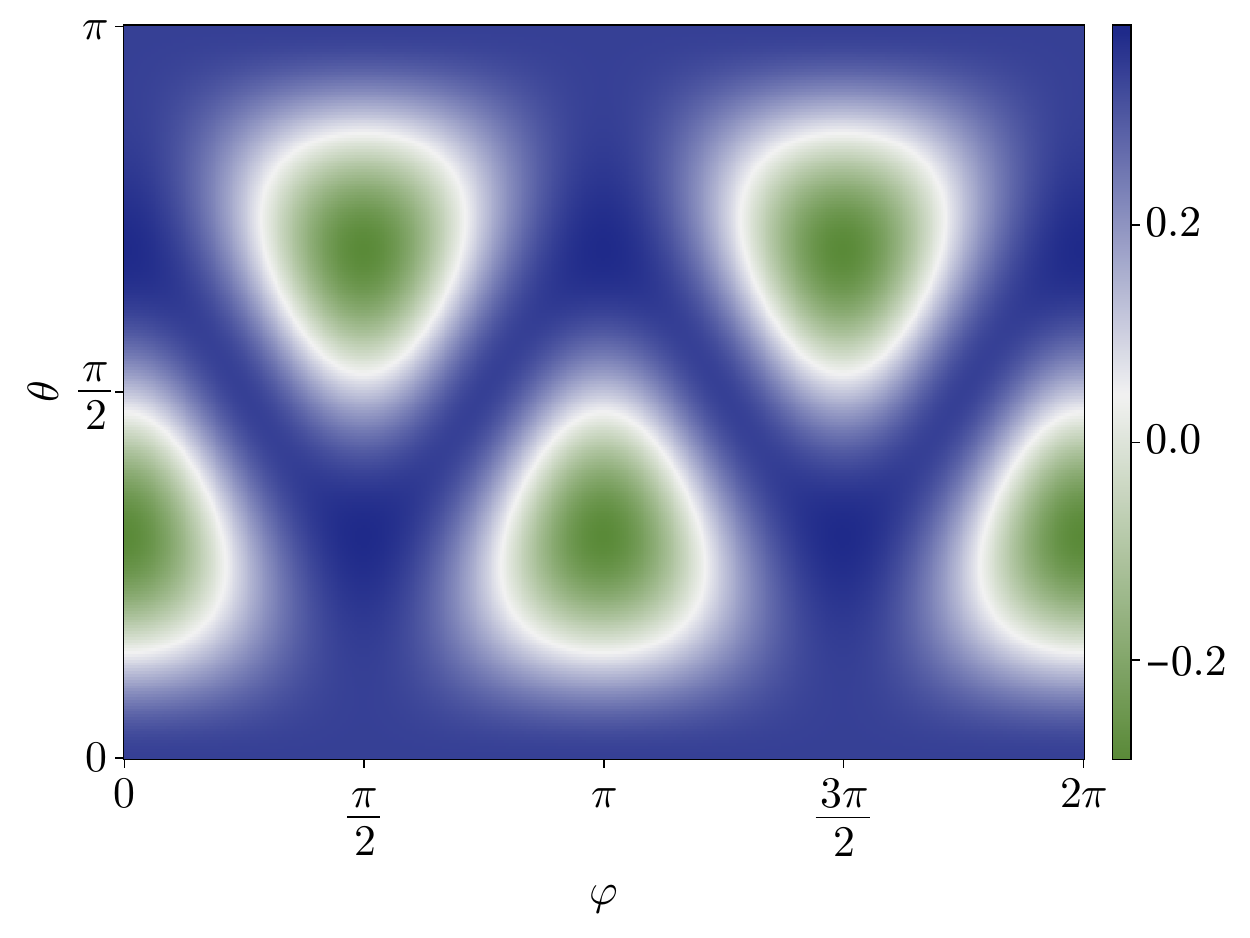}
    \caption{Steady state Wigner function for $N=5$ showing the preparation of a nonclassical steady state with the tetrahedral symmetry for $H = H_\mathrm{tetra}$, coupling operators $L_\mathrm{tetra}$ and bosonic AS of frequency $\omega$. Parameters: $h =\omega$, $g = \sqrt{\gamma \kappa/(2N)}$ with $\gamma = \kappa = \omega/2$. }
    \label{fig::N5}
\end{figure}
\section{Comparison with Hamiltonian dynamics}
In this section, we elaborate on the intrinsic nature of the steady states that we prepare. As stated in the main text, there exists no pure state for $N=5$ which is $2$-anticoherent. Nevertheless, our general protocol still enables the preparation of steady states with tetrahedral symmetry (and thus $2$-anticoherent, see previous section). This is possible because the steady states are mixed rather than pure. As illustrated in Fig.~\ref{fig::N5}, in contrast to pure Hamiltonian dynamics, our method yields steady states with the tetrahedral symmetry that can remain highly nonclassical.

Finally, we stress an important difference with Hamiltonian engineering. A Hamiltonian with the target symmetry does not generally possess a unique ground state. In that case, symmetry only guarantees that the ground-state manifold is invariant under the group action, not that an individual pure ground states inherit the symmetry. As a simple illustration, in the $\mathbb{D}_2$ case the Hamiltonian $H_S=(h/N)S_z^2$ is $\mathbb{D}_2$-symmetric, yet for odd $N$ and $h>0$ its ground state is twofold degenerate, spanned by $\ket{N/2,\pm 1/2}$. Individual ground states in this manifold need not be $\mathbb{D}_2$-invariant. 
By contrast, in our dissipative construction, the steady state is unique and symmetry inheritance applies directly to the prepared state itself. We note that this simple observation generalizes to higher symmetry groups.
\bibliographystyle{apsrev4-2}
\bibliography{bib}